\documentclass[10pt,twocolumn,letterpaper]{article}

\usepackage{wacv}              

\usepackage{graphicx}
\usepackage{amsmath}
\usepackage{amssymb}
\usepackage{booktabs}
\usepackage{fancyhdr}

\fancypagestyle{firstpagefooter}{
  \fancyhf{} 
  \fancyfoot[L]{\small *\textit{Submitted}} 
}

\usepackage[pagebackref,breaklinks,colorlinks]{hyperref}

\usepackage[capitalize]{cleveref}
\crefname{section}{Sec.}{Secs.}
\Crefname{section}{Section}{Sections}
\Crefname{table}{Table}{Tables}
\crefname{table}{Tab.}{Tabs.}

\def\wacvPaperID{*****} 
\def\confName{WACV}
\def\confYear{2025}

\begin{document}

\title{\textit{NestedMorph}: Enhancing Deformable Medical Image Registration with Nested Attention Mechanisms}

\setlength{\parskip}{0pt}
\setlength{\parsep}{0pt}
\setlength{\topsep}{0pt}
\setlength{\partopsep}{0pt}

\author{Gurucharan Marthi Krishna Kumar\\
Montreal Neurological Institute, McGill University\\
{\tt\small gurucharan.marthikrishnakumar@mail.mcgill.ca}
\and
Janine Mendola\\
Dept. of Ophthalmology, McGill University\\
{\tt\small janine.mendola@mcgill.ca}
\and
Amir Shmuel\\
Montreal Neurological Institute, McGill University\\
{\tt\small amir.shmuel@mcgill.ca}
}

\maketitle
\thispagestyle{firstpagefooter}

\begin{abstract}
   Deformable image registration is crucial for aligning medical images in a nonlinear fashion across different modalities, allowing for precise spatial correspondence between varying anatomical structures. This paper presents NestedMorph, a novel network utilizing a Nested Attention Fusion approach to improve intra-subject deformable registration between T1-weighted (T1w) MRI and diffusion MRI (dMRI) data. NestedMorph integrates high-resolution spatial details from an encoder with semantic information from a decoder using a multi-scale framework, enhancing both local and global feature extraction. Our model notably outperforms existing methods, including CNN-based approaches like VoxelMorph, MIDIR, and CycleMorph, as well as Transformer-based models such as TransMorph and ViT-V-Net, and traditional techniques like NiftyReg and SyN. Evaluations using the HCP dataset demonstrate that NestedMorph achieves superior performance across key metrics, including SSIM, HD95, and SDlogJ, with the highest SSIM of 0.89, the lowest HD95 of 2.5 and SDlogJ of 0.22. These results highlight NestedMorph's ability to capture both local and global image features effectively, leading to superior registration performance. The promising outcomes of this study underscore NestedMorph's potential to significantly advance deformable medical image registration, providing a robust framework for future research and clinical applications. The source code and our implementation are available at: \href{https://github.com/AS-Lab/Marthi-et-al-2024-NestedMorph-Deformable-Medical-Image-Registration}{https://github.com/AS-Lab/Marthi-et-al-2024-NestedMorph-Deformable-Medical-Image-Registration}

\end{abstract}

\section{Introduction}
\label{sec:intro}

Deformable image registration is essential for medical imaging tasks as it aligns fixed and moving images by finding a nonlinear mapping that minimizes their differences. This process establishes spatial correspondence through a continuous deformation field, which can be modeled in various ways, leading to numerous methods. Recent advancements in deformable image registration have been significantly enhanced by convolutional neural networks (CNNs), which have markedly improved both inference speed and accuracy compared to traditional methods. The field has seen a rapid rise in deep learning-based approaches. Recently, Vision Transformers (ViTs) have been integrated into deformable image registration to overcome the limitations of restricted receptive fields typical in CNN-based methods. However, Transformer-based techniques still face challenges in accurately modeling deformation fields. Here, we present a novel unsupervised hybrid CNN-Transformer based deep learning method for intra-subject registration, ensuring seamless alignment between T1w and dMRI templates for accurate anatomical and diffusion data correspondence. 

dMRI of the brain induces distortions and partial volume effects from cerebrospinal fluid, necessitating a nonlinear alignment approach. While signal averaging helps mitigate these effects, some studies opt for rigid body affine registration, which neglects nonlinear distortion \cite{carrozzi2023methods}. To address this challenge, recent advancements in high-resolution dMRI have introduced nonlinear algorithms to improve alignment. Deep learning methods have also evolved, leveraging extensive datasets of paired or unpaired images to enhance deformable medical image registration, departing from traditional approaches \cite{kim2019incorporating}.

Current research in medical image registration focuses on applying registration networks to diverse 3D images from public datasets but lacks comprehensive evaluation for intra-subject deformable registration between T1w and dMRI scans. While recent deep learning techniques such as VoxelMorph \cite{balakrishnan2019voxelmorph} and CycleMorph \cite{kim2021cyclemorph} leverage CNNs for deformation fields, they struggle with long-range spatial relations \cite{chen2022transmorph}. In contrast, Transformer-based networks such as TransMorph \cite{chen2022transmorph} and ViT-V-Net \cite{chen2021vit} excel in understanding spatial correspondence, making them well-suited for intra-subject deformable registration between T1w and diffusion MRI scans \cite{li2021medical}. Our proposed NestedMorph network is designed to improve intra-subject deformable registration between T1-weighted and dMRI modalities, and it includes a comparative analysis with state-of-the-art architectures as well as the proposed network itself. The key contributions of this study are summarized as follows

\begin{itemize}
    \item We present a Nested Attention Fusion network that integrates an encoder's high-resolution details with a decoder's semantic information, improving the alignment and accuracy of deformable registration.
    
    \item Our method enhances deformation field estimation through continuous deformation information flow and a multi-scale framework that combines low- and high-resolution processing for a precise representation.
    
    \item We compare our unsupervised method with leading CNN and Transformer-based approaches using the HCP dataset, showing superior performance in intra-subject deformable registration between dMRI and T1-weighted MRI data.
\end{itemize}

\section{Related Works}
Deformable Image Registration methods are generally divided into two main types: supervised and unsupervised approaches. Supervised methods benefit from utilizing external data, such as label maps, to guide the training process. On the other hand, unsupervised techniques aim to perform registration by exploring the inherent characteristics of the data itself. Supervised methods \cite{cao2017deformable,sokooti2017nonrigid,rohe2017svf,yang2017quicksilver}, require ground-truth deformation fields, which can be computationally expensive to obtain. As a result, research has increasingly shifted towards unsupervised methods, which do not rely on ground-truth deformation fields. 

Among the unsupervised methods, an unsupervised learning method for image registration that leverages a low-dimensional stochastic parameterization of deformation was proposed \cite{krebs2018unsupervised} . VoxelMorph \cite{balakrishnan2019voxelmorph} utilizes a CNN-based encoder-decoder architecture akin to UNet for computing deformation fields between moving and fixed image pairs, with a Spatial Transformer network warping the moving image iteratively. MIDIR \cite{qiu2021learning} employs B-spline parameterization to generate diffeomorphic deformation fields, utilizing Mutual Information loss and a stationary velocity field (SVF) for smooth deformation. CycleMorph \cite{kim2021cyclemorph} incorporates two registration networks to compute the displacement vector fields and introduces a cycle-consistent learning model for unsupervised registration while maintaining topology between images. 

Within transformer-based methods, TransMorph \cite{chen2022transmorph} presents a novel hybrid model combining Transformers and CNNs specifically for 3D medical image registration. This model leverages the self-attention mechanism and the extensive effective receptive field of Transformers to enhance registration accuracy. ViT-V-Net \cite{chen2021vit} uses a hybrid CNN-Transformer architecture for self-supervised volumetric image registration. It applies the Vision Transformer (ViT) to high-level image features to capture long-range dependencies, while long skip connections maintain localization information between the encoder and decoder stages. The self-attention mechanism in Transformers overcomes some of the constraints of CNNs, especially in capturing long-range dependencies within the deformation field. This need underscores the importance of adequate patch representation to fully leverage long-range information in Transformer-based models.

\section{Proposed Method}
\label{sec:methods}
\begin{figure*}
    \centering
    \includegraphics[scale=0.075]{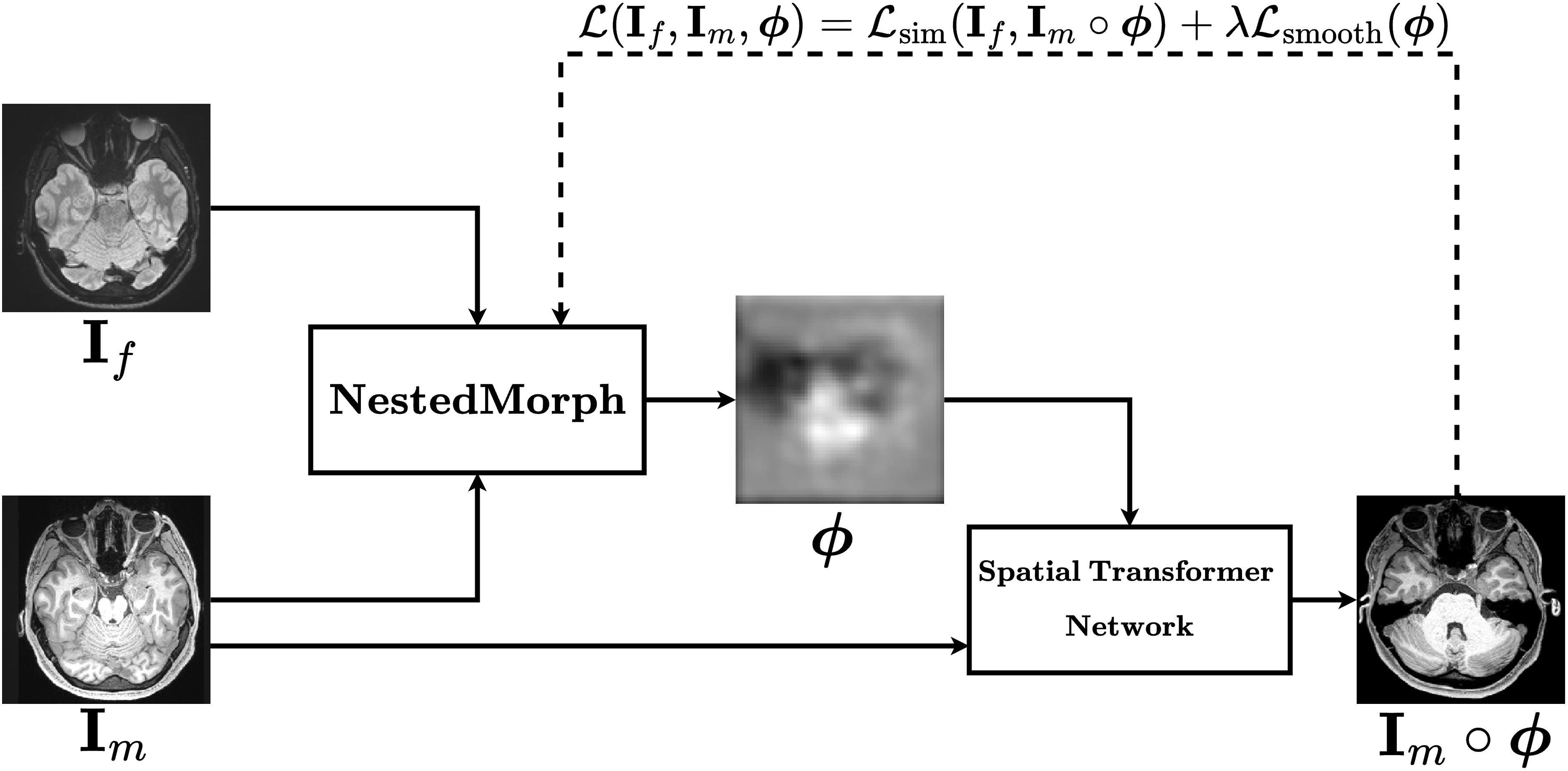}
    \caption{\textbf{Overall Framework of the Deformable Registration.} This figure presents the complete NestedMorph registration pipeline, where the fixed and moving images are aligned using the NestedMorph network and a spatial transformer network to generate a refined deformation field. The architecture integrates both local and global information, improving registration accuracy for medical image modalities such as T1-weighted MRI and dMRI.}
    \label{fig:framework}
  \hfill
\end{figure*}

Let \( I_f \) and \( I_m \) denote the fixed and moving images, which are the mean dMRI and T1w images, respectively. Let \( \phi \) represent the deformation field that aligns the two images. The images \( I_f \) and \( I_m \) are concatenated and fed into the NestedMorph model. The goal here is to produce a deformation field \( \phi \) that maps the grid of the moving image \( I_m \) to the grid of the fixed image \( I_f \), ensuring accurate alignment between the two. The encoder architecture employs a multi-level design that combines overlapping patch embeddings with dual attention mechanisms to extract and process features from 3D medical images. Initially, the input image is divided into overlapping patches and embedded into a lower-dimensional space through convolution operations. This embedding process reduces spatial resolution while preserving crucial information. The embedded patches are then processed through multiple stages, where efficient attention and channel attention mechanisms enhance the model's ability to focus on both local and global features simultaneously. At each stage, the spatial dimensions are further reduced, and the feature richness is increased. The processed features are normalized and reshaped, resulting in a series of multi-scale feature maps. The decoder employs a Dual Attention Enhanced Transformer (DAE-Former) blocks \cite{azad2023dae} to maintain long-range dependencies and structural integrity. The final layers incorporate a Large Kernel Attention (LKA) \cite{azad2024beyond} module to handle local and global features, crucial for high-resolution image processing. The last layer outputs a 3-channel deformation field, used by the spatial transformer network (STN) to achieve precise image registration, ensuring alignment between \( I_f \) and \( I_m \). Figure \ref{fig:framework} presents the overall framework of the deformable registration pipeline

\begin{figure*}
    \centering
    \includegraphics[scale=0.0765]{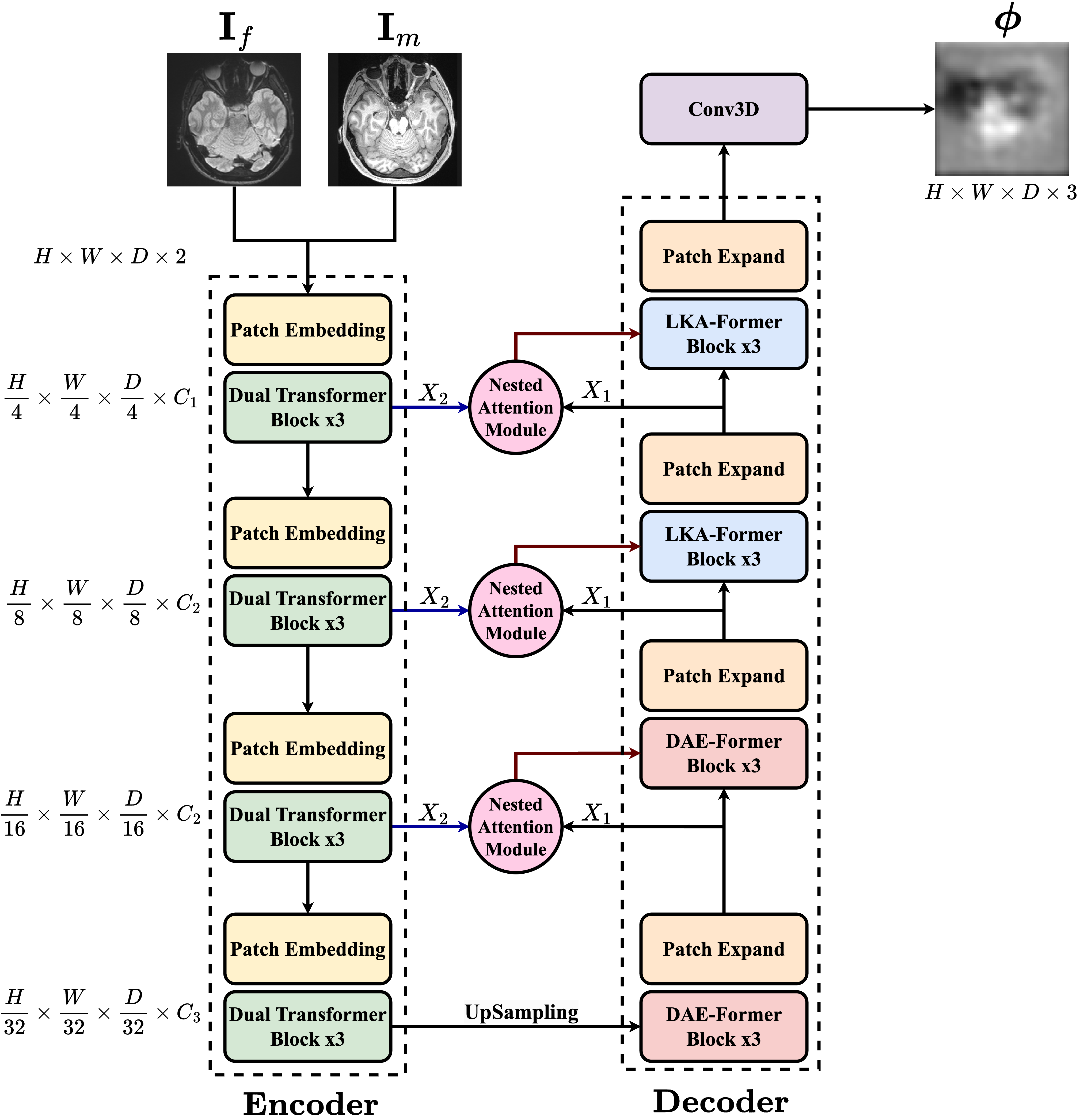}
    \caption{\textbf{Nested Morph Architecture.} The NestedMorph architecture leverages dual attention mechanisms to fuse encoder and decoder features for accurate alignment between fixed and moving medical images. The overall architecture, shown here, highlights how Nested Attention Modules contribute to precise deformation field estimation.}
    \label{fig:nestedmorph}
  \hfill
\end{figure*}

\subsection{Encoder}
Figure \ref{fig:nestedmorph} illustrates the NestedMorph architecture. The NestedMorph model features an encoder block that integrates four encoder layers with dual Transformer mechanisms and a patch embedding module. Given the concatenated image of size \( H \times W \times D \times 2 \), the encoder processes the image through three stages. In the first stage, the image is divided into overlapping patches with a stride of 4, reducing the spatial dimensions to \( \frac{H}{4} \times \frac{W}{4} \times \frac{D}{4} \times C_1 \) channels. The output is then processed by the dual Transformer layers of the first encoder block, which apply Efficient Attention and Channel Attention mechanisms to capture both local and global dependencies.

In the subsequent stages, patches with a stride of 2 further reduce the spatial dimensions while doubling the number of channels. The feature maps from each stage are processed by the dual Transformer layers of the corresponding encoder blocks. This progression results in a final output size of \( \frac{H}{32} \times \frac{W}{32} \times \frac{D}{32} \times C_3 \). Throughout these stages, patch merging operations aggregate information from smaller patches, enabling the model to capture multi-scale features essential for precise deformable registration of medical images.

\subsubsection{Efficient and Channel Attention}
Efficient attention \cite{shen2021efficient}, enhances traditional self-attention by optimizing the computation of context matrices. It is defined as:

\begin{equation}
E(\mathbf{Q}, \mathbf{K}, \mathbf{V}) = \rho_q(\mathbf{Q}) \left( \rho_k(\mathbf{K})^T \mathbf{V} \right)
\end{equation}

where \(\rho_q\) and \(\rho_k\) are softmax normalization functions for queries and keys. Unlike dot-product attention, efficient attention bypasses the direct calculation of pairwise similarities. Instead, it employs keys represented as single-channel feature maps, which serve as weightings for all positions. These weightings are then used to aggregate value features through weighted summation, thereby constructing a global context vector. These global attention maps capture semantic aspects of the entire input feature rather than just positional similarities. This approach reduces computational complexity while preserving high representational power.

Channel attention, also known as transpose attention \cite{ali2021xcit}, is designed to handle larger input sizes efficiently by focusing on the entire channel dimension. The mechanism is defined as:

\begin{equation}
C(\mathbf{Q}, \mathbf{K}, \mathbf{V}) = \mathbf{V} \cdot \textit{Softmax}\left(\frac{\mathbf{K}^T \mathbf{Q}}{\tau}\right)
\end{equation}

In this formulation, the keys and queries are transposed, and the attention weights are computed based on the cross-covariance matrix. The temperature parameter \(\tau\) normalizes the queries and keys, adjusting the scale of the inner products prior to applying the Softmax function. This adjustment helps control the sharpness or uniformity of the resulting attention weights.

\subsubsection{Dual Attention Transformer Block}

A previous study \cite{guo2022attention} demonstrates that integrating efficient and channel attention mechanisms improves a model's ability to capture richer contextual features compared to using a single attention type. Consequently, we design a dual Transformer block that incorporates both channel attention and efficient attention. Building on a previous work \cite{azad2023dae}, we implement this dual attention block to enhance feature extraction and representation.

The dual attention transformer block integrates two main components: an Efficient Attention mechanism followed by an Add \& Norm step, and a Transpose Attention block that performs Channel Attention, succeeded by another Add \& Norm step.

\begin{align}
\textit{EA}_{b}(\mathbf{X}, \mathbf{Q_1}, \mathbf{K_1}, \mathbf{V_1}) &= \textit{EA}(\mathbf{Q_1}, \mathbf{K_1}, \mathbf{V_1}) + \mathbf{X} \\
\textit{M1} &= \textit{MLP}(\textit{LN}(\textit{EA}_{b})) \\
\textit{CA}_{b}(\textit{EA}_{b}, \mathbf{Q_2}, \mathbf{K_2}, \mathbf{V_2}) &= \textit{CA}(\textit{EA}_{b} + \textit{M1}) + \textit{M1} \\
\textit{M2} &= \textit{MLP}(\textit{LN}(\textit{CA}_{b})) \\
\textit{DA} &= \textit{CA}_{b} + \textit{M2}
\end{align}

In this context, \textit{EA} and \textit{CA} denote the Efficient Attention and Channel Attention mechanisms, respectively. \( \textit{CA}_{b} \) refers to the Channel Attention block, and \( \textit{EA}_{b} \) denotes the Efficient Attention block. The queries, keys, and values used in Efficient Attention are \( \mathbf{Q_1} \), \( \mathbf{K_1} \), and \( \mathbf{V_1} \), derived from the input feature \( \mathbf{X} \), while \( \mathbf{Q_2} \), \( \mathbf{K_2} \), and \( \mathbf{V_2} \) are used for the Channel Attention block. Additionally, \( \textit{LN} \) refers to Layer Normalization, applied to stabilize and speed up training by normalizing the outputs of the attention and \( \textit{MLP} \) operations. The Mix-FFN \cite{huang2021missformer} feed-forward network, denoted as \( \textit{MLP} \), is defined as:

\begin{equation}
\textit{MLP}(X) = \textit{FC}(\textit{GELU}(\textit{DW-Conv}(\textit{FC}(\mathbf{X}))))
\end{equation}

where \(\textit{FC}\) represents a fully connected layer, \(\textit{GELU}\) is the GELU activation function \cite{hendrycks2016gaussian} and \(\textit{DW-Conv}\) denotes depth-wise convolution.

\subsection{Decoder}
In the decoder, the initial layers employ Dual Attention Enhanced Transformer (DAE-Former) blocks \cite{azad2023dae} to preserve long-range dependencies within lower-resolution images. These blocks are adept at integrating both spatial and channel attention, ensuring that the image’s structural integrity is maintained while efficiently handling complex spatial interactions. The final layers of the decoder incorporate the Large Kernel Attention (LKA) \cite{azad2024beyond} module, which excels in managing local and global features. This module addresses challenges associated with high-resolution image processing in transformers by enhancing the model's ability to capture fine details and intricate patterns. Additionally, we employ a Nested Attention module that leverages skip connections to integrate features from various layers, enabling each decoder layer to access fine-grained spatial details crucial for generating accurate output masks. Specifically, we utilize cross-attention where the encoder’s output, denoted as \(\mathbf{X_2}\), serves as the query input. This allows the decoder to focus on high-level features from the encoder. Meanwhile, the keys and values are derived from the output of the lower decoder layer, \(\mathbf{X_1}\). This setup enables the decoder to successfully attend to and integrate both the detailed spatial information and the high-level contextual cues. The last layer outputs a deformation field \( \phi \) with 3 channels, corresponding to the three spatial directions. This deformation field is then applied to a spatial transformer network for deformable registration, ensuring alignment between the moving \( I_m \) and fixed \( I_f \) images

\subsubsection{Nested Attention Fusion Module}

\begin{figure}
    \centering
    \includegraphics[scale=0.065]{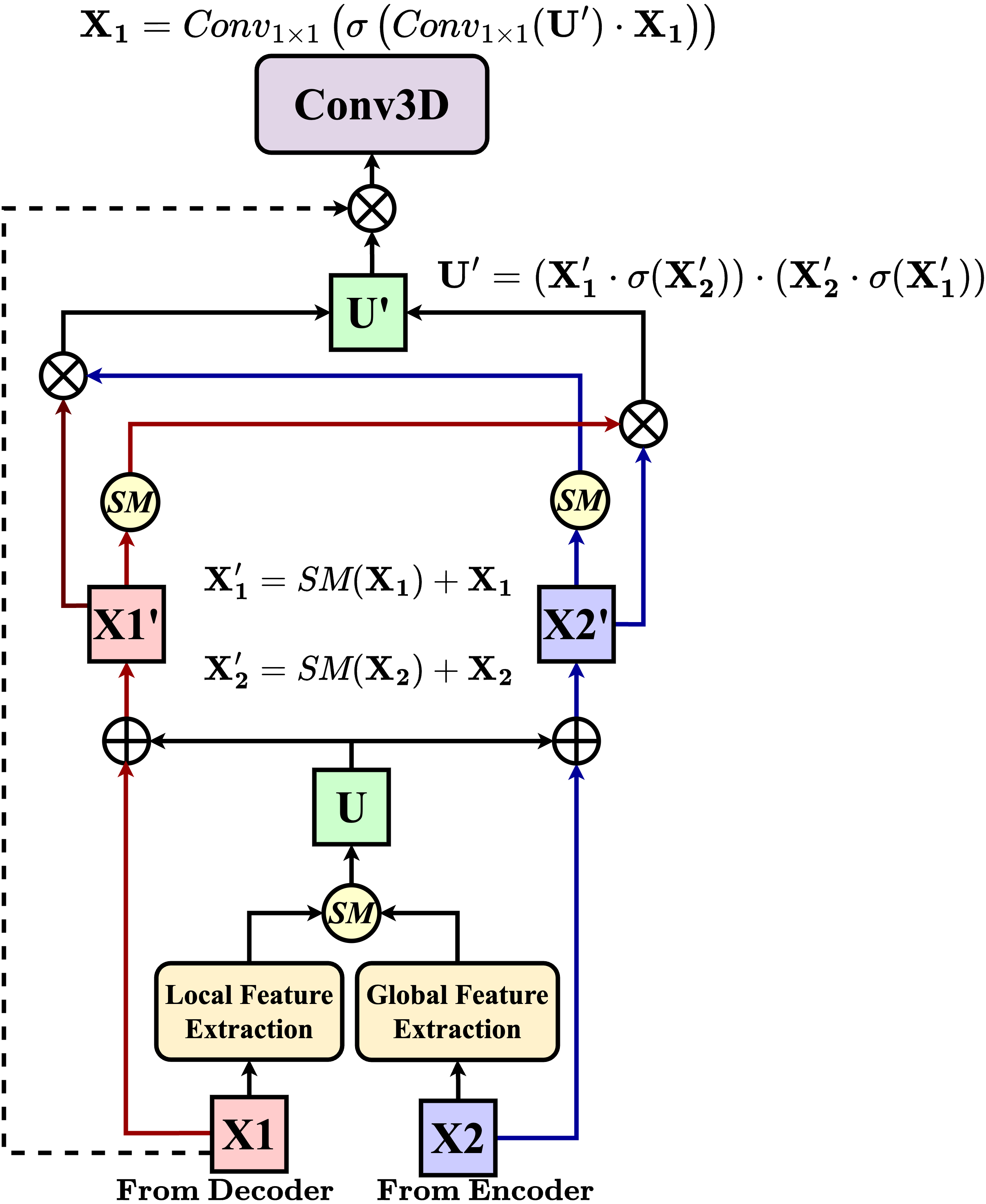}
    \caption{\textbf{Nested Attention Fusion Module.} The Nested Attention Module fuses local and global features extracted from both encoder and decoder outputs through a combination of spatial modulations and attention mechanisms. This design effectively captures both fine-grained local details and long-range dependencies, enhancing the deformable image registration process.}
    \label{fig:nestedattention}
    \hfill
\end{figure}

The Nested Attention Fusion Module, adapted from \cite{kolahi2024msa2net}, processes decoder and encoder outputs (\(\mathbf{X_1}\) and \(\mathbf{X_2}\), respectively). The module utilizes multi-scale fusion to aggregate \(\mathbf{X_1}\) and \(\mathbf{X_2}\) in a semantically coherent manner, preparing the fused feature map for further processing. Spatial selection dynamically adjusts receptive fields to emphasize crucial features, while spatial interaction and cross modulation enhance the feature maps with both local details and global context. Figure \ref{fig:nestedattention} shows the Nested Attention Fusion Module.

The module first employs Global Extraction to capture broad, global context from the encoder features (\(\mathbf{X_2}\)) using average and max pooling, which are then projected into a refined feature space. Simultaneously, Feature Extraction enhances the local spatial details of the decoder features (\(\mathbf{X_1}\)) through depthwise separable and dilated convolutions, followed by dimensionality reduction. These processed local and global features are then combined in the Multi-Scale Fusion stage, which integrates them through summation and normalization to get \(\mathbf{U}\). The resulting fused feature map is passed through a Nested Gated Attention mechanism, which applies a gated attention technique to calculate selective weights using a channel-wise softmax function, leading to spatially selected versions \( \mathbf{X_1}' \) and \( \mathbf{X_2}' \).

\begin{equation}
\textit{SM} = \text{Softmax}(\text{Conv}_{1 \times 1}(\mathbf{U}))
\end{equation}

\begin{equation}
\mathbf{X_1'} = \textit{SM}(\mathbf{X_1}) + \mathbf{X_1}, \quad \mathbf{X_2'} = \textit{SM}(\mathbf{X_2}) + \mathbf{X_2}
\end{equation}

In the next stage, \( \mathbf{X_1}' \), with its refined local details, is enriched by global contexts from \( \mathbf{X_2}' \) through spatial weights applied via a sigmoid function, resulting in \( \mathbf{X_1}'' \). Initially representing broader contexts with applied weights, it integrates detailed contexts from \( \mathbf{X_1}' \), evolving into \( \mathbf{X_2}'' \). This is further multiplied to get a fused feature map \( \mathbf{U'} \):

\begin{equation}
\mathbf{U'} = ( \mathbf{X_1'} \cdot \sigma(\mathbf{X_2'}) ) \cdot ( \mathbf{X_2'} \cdot \sigma(\mathbf{X_1'}) )
\end{equation}

where, \(\sigma(\mathbf{X_1'})\) and \(\sigma(\mathbf{X_2'})\) refer to local and global spatial weights, respectively.

Finally, this fused feature map \( \mathbf{U'} \) is projected back onto the initial input from the decoder \( \mathbf{X_1'} \), resulting in an attention mechanism that ensures both detailed spatial information and high-level contextual cues are integrated. This leads to a robust and contextually informed representation, improving output accuracy.

\begin{equation}
\mathbf{X_1} = \textit{Conv}_{1 \times 1} \left( \sigma \left( \textit{Conv}_{1 \times 1} (\mathbf{U'}) \cdot \mathbf{X_1} \right) \right)
\end{equation}

This attention mechanism ensures that both the detailed spatial information and the high-level contextual cues are effectively integrated, resulting in a contextually informed representation for improved output accuracy.

\subsection{Spatial Transformer Network}
The \(STN\)\cite{jaderberg2015spatial} warps the moving image \(I_m\) with the predicted deformation field \(\phi\), resulting in the deformed image (\(I_m \circ \phi\)).  For each voxel \( v \), we compute \( v_0 = v + u(v) \) in \( I_m \) by performing trilinear interpolation using the values of the eight neighboring voxels \cite{balakrishnan2019voxelmorph}.
\[
I_m \circ \phi(v) = \sum_{u \in Z(v_0)} I_m(u) \prod_{d \in \{x, y, z\}} \left(1 - |v_0^d - u^d|\right),
\]
where \( Z(v_0) \) denotes the set of neighboring voxels surrounding \( v_0 \), and \( d \) spans across the dimensions of $I_m$. This network minimizes differences between \(I_m \circ \phi\) and \(I_f\) through iterative training, resulting in a registered deformed image.

\section{Experimental Results}
\subsection{Dataset and Preprocessing}
This study utilized data from 205 subjects from the HCP Aging Dataset \cite{glasser2013minimal,van2013wu}, each contributing T1w and dMRI scans. Of the 205 subjects, 180 were used for training the network, and the remaining 25 were reserved for testing. T1w scans had a voxel size of 0.7 mm isotropic, while dMRI scans had a voxel size of 1.25 mm isotropic, with single-shell dMRI data at $b=1000 s/\text{mm}^2$ in 90 directions and 18 b0 images. The dMRI data were averaged across all 108 volumes to create 3D mean dMRI images \cite{chen2019diffusion}. Both T1w and mean dMRI scans were resized to 128×128×128 voxels. The T1w images were then affinely registered to the mean dMRI using SimpleITK in Python, resulting in pairs of moving T1w and fixed mean dMRI images. These 180 training pairs were split, with 80\% (144) used for model training and 20\% (36) for validation.

\subsection{Training Parameters}
The NestedMorph model was implemented in Python using the PyTorch framework on an NVIDIA A100 GPU with 40 GB of memory. The training was conducted with a batch size of 4 using stochastic gradient descent, with a base learning rate of \(1 \times 10^{-4}\) and a weight decay of \(3 \times 10^{-5}\). The model was trained for 100 epochs using a composite loss function that computes the similarity between the deformed moving image and the fixed image, and imposes smoothness on the deformation field to regularize it \cite{chen2020generating}. 

The composite loss function \( \mathcal{L} \) is defined as:

\begin{equation}
\mathcal{L}(\mathbf{I}_f, \mathbf{I}_m, \boldsymbol{\phi}) = \mathcal{L}_{\text{sim}}(\mathbf{I}_f, \mathbf{I}_m \circ \boldsymbol{\phi}) + \lambda \mathcal{L}_{\text{smooth}}(\boldsymbol{\phi})
\end{equation}

Here, \( \mathcal{L}_{\text{sim}} \) is calculated using Normalized Cross-Correlation (NCC), while \( \mathcal{L}_{\text{smooth}} \) employs an isotropic diffusion term, similar to Gaussian smoothing:

\[
\mathcal{L}_{\text{sim}}(\mathbf{I}_f, \mathbf{I}_m \circ \boldsymbol{\phi}) = 1 - \text{NCC}(\mathbf{I}_f, \mathbf{I}_m \circ \boldsymbol{\phi})
\]

\[
\mathcal{L}_{\text{smooth}}(\boldsymbol{\phi}) = \|\nabla \boldsymbol{\phi}\|^2
\]

where \( \nabla \boldsymbol{\phi} \) represents the gradient of the deformation field \( \boldsymbol{\phi} \).

\subsection{Baselines and comparison}
We compare the proposed NestedMorph network with state-of-the-art unsupervised deformable registration techniques, including both CNN-based methods such as VoxelMorph \cite{balakrishnan2019voxelmorph}, MIDIR \cite{qiu2021learning}, and CycleMorph \cite{kim2021cyclemorph}, and transformer-based methods like TransMorph \cite{chen2022transmorph} and ViT-V-Net \cite{chen2021vit}. We also evaluate traditional methods including NiftyReg \cite{modat2010fast} and SyN \cite{avants2008symmetric}. To assess registration accuracy, we use established similarity metrics such as the Structural Similarity Index (SSIM) \cite{wang2004image}, the 95th percentile Hausdorff distance (HD95) \cite{huttenlocher1993comparing}, and the standard deviation of the logarithm of the Jacobian determinant (SDlogJ) \cite{liu2024finite} to compare the similarity of the deformed image \( \mathbf{I}_m \circ \boldsymbol{\phi} \) to the fixed image \( \mathbf{I}_f \).

\subsection{Quantitative Results}
As outlined above, 25 subjects from the dataset were reserved for testing and used to compare the performance of different registration networks. For each subject, the T1w scans were initially affinely registered to their dMRI scans. Subsequently, various registration networks were deployed to perform deformable registration on the registered images.

From Table \ref{tab:task_comparison}, it is evident that the proposed NestedMorph model outperforms all other models across the three evaluated metrics: SSIM, HD95, and SDlogJ. Specifically, NestedMorph achieves the highest SSIM of 0.89, indicating superior structural similarity and image quality after registration. It also demonstrates the lowest HD95 of 2.5 and SDlogJ of 0.22, reflecting better boundary alignment and reduced deformation irregularities, respectively. While models like TransMorph and CycleMorph show strong performance, with TransMorph coming close to NestedMorph in SSIM and having slightly higher HD95 and SDlogJ values, they do not surpass the NestedMorph model. Similarly, while NiftyReg and SyN improve upon the Affine method, they remain outperformed by the more advanced data-driven models. This comparison underscores the efficacy of integrating advanced modules into our proposed NestedMorph, thus enhancing its ability to deliver high-quality registrations with improved precision and reliability in medical images.

\begin{table}[ht]
    \centering
    {\small{
    \begin{tabular}{@{}l@{\hskip 0.05in}ccc@{}}
        \toprule
        \textbf{Model} & \textbf{SSIM $\uparrow$} & \textbf{HD95 $\downarrow$} & \textbf{SDlogJ $\downarrow$} \\
        \midrule
        Affine & 0.72 ± 0.05 & 5.1 ± 0.6 & 0.42 ± 0.04 \\
        NiftyReg & 0.76 ± 0.07 & 4.7 ± 0.7 & 0.35 ± 0.04 \\
        SyN & 0.75 ± 0.06 & 4.9 ± 0.8 & 0.38 ± 0.05 \\
        MIDIR & 0.79 ± 0.08 & 3.8 ± 0.5 & 0.30 ± 0.03 \\
        VoxelMorph & 0.82 ± 0.07 & 3.6 ± 0.4 & 0.28 ± 0.02 \\
        CycleMorph & 0.85 ± 0.06 & 3.2 ± 0.3 & 0.27 ± 0.03 \\
        TransMorph & 0.88 ± 0.07 & 2.8 ± 0.4 & 0.25 ± 0.02 \\
        ViT-V-Net & 0.84 ± 0.05 & 3.0 ± 0.3 & 0.26 ± 0.03 \\
        \textbf{Proposed NestedMorph} & \textbf{0.89 ± 0.06} & \textbf{2.5 ± 0.3} & \textbf{0.22 ± 0.02} \\
        \bottomrule
    \end{tabular}
    }}
    \caption{Comparison of models based on SSIM, HD95, and SDlogJ metrics. The values are followed by standard deviations. The highest metrics are highlighted in bold.}
    \label{tab:task_comparison}
\end{table}

\begin{figure*}
    \centering
    \includegraphics[scale=0.0525]{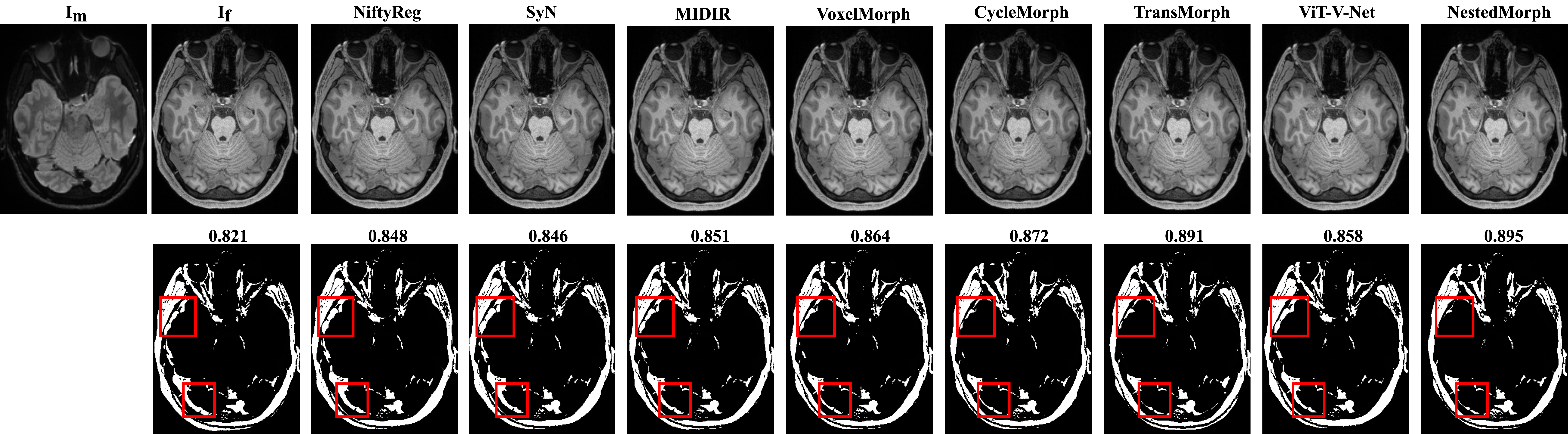}
    \caption{\textbf{Comparison of registration methods applied to medical images.} The first two columns show the moving ($I_m$) and the fixed ($I_f$) image. Subsequent columns display registered images (top row) and their corresponding binary difference images (bottom row) for various methods. The SSIM values above the binary images quantify the similarity between the registered image and $I_f$. NestedMorph achieves the highest SSIM, indicating the best registration performance. Red boxes highlight key differences in the binary images.}
    \label{fig:registration_comparison}
  \hfill
\end{figure*}

Figure \ref{fig:registration_comparison} presents a comparative analysis of several registration methods applied to medical images. The moving image ($I_m$) and the fixed image ($I_f$) are shown alongside the registered images generated by different methods. The top row displays the registered images, while the bottom row shows the corresponding binary difference images, which highlight the discrepancies between the registered and fixed images. The red boxes in the binary difference images highlight areas where the registration performance is easily noticeable. Each method's performance is quantitatively assessed using SSIM, which is displayed above the binary difference images. The Affine method, although foundational, shows a moderate SSIM, reflecting its basic approach to alignment. NiftyReg and SyN show slight improvements in SSIM compared to the Affine method, but they still fall short of the performance achieved by deep learning-based approaches. In contrast, advanced techniques such as MIDIR and VoxelMorph and CycleMorph improve the SSIM slightly, indicating better alignment and image fidelity. Overall, TransMorph and the proposed NestedMorph method achieve the highest SSIM scores, with NestedMorph slightly outperforming all others, signifying its superior performance in accurately aligning the moving image to the fixed image. 

\subsection{Ablation study on the NestedMorph model}
To evaluate the impact of various network modules on registration performance, we conducted an ablation study using the testing dataset. In Section \ref{sec:model_capacity}, we investigated the impact of various parameters on performance, including the number of attention heads, network depth, patch size, and batch size. Additionally, to assess how different network modules affect registration performance, we performed an ablation study in Section \ref{sec:attention}. This study explored various configurations of attention mechanisms and the number of DAE-Former and LKA blocks.

\subsubsection{Model Capacity and Fine Tuning}
\label{sec:model_capacity}

Table \ref{tab:parameter_performance} presents a comparison of various parameters, namely batch size, number of heads, patch size, and number of layers, to evaluate their impact on model performance in terms of SSIM score and HD95. The results suggest that certain configurations lead to better performance, providing insights into optimizing model architecture for improved accuracy and robustness.

\begin{table}[ht]
    \centering
    \begin{tabular}{@{}lcc@{}}
        \toprule
        \textbf{Parameter} & \textbf{SSIM $\uparrow$} & \textbf{HD95 $\downarrow$} \\
        \midrule
        \textit{Batch Size} & & \\
        2 & 0.812±0.072 & 2.7±0.4 \\
        8 & 0.825±0.069 & 2.5±0.3 \\
        \midrule
        \textit{Number of Heads} & & \\
        1 & 0.818±0.071 & 2.8±0.4 \\
        8 & 0.832±0.068 & 2.6±0.3 \\
        \midrule
        \textit{Patch Size} & & \\
        \(3 \times 3 \times 3 \) & 0.835±0.070 & 2.4±0.3 \\
        \(6 \times 6 \times 6 \) & 0.827±0.072 & 2.6±0.4 \\
        \midrule
        \textit{Number of Layers} & & \\
        4 & 0.830±0.073 & 2.5±0.4 \\
        8 & 0.822±0.069 & 2.7±0.3 \\
        \bottomrule
    \end{tabular}
    \caption{Impact of Various Parameters on Model Performance}
    \label{tab:parameter_performance}
\end{table}

\begin{table*}[ht]
    \centering
    \begin{tabular}{lccccccc@{}}
        \toprule
        \textbf{Method} & \textbf{Params (M)} & \textbf{EA} & \textbf{CA} & \textbf{DAE-Former (\#)} & \textbf{LKA (\#)} & \textbf{SSIM $\uparrow$} & \textbf{HD95} $\downarrow$ \\
        \midrule
        NestedMorph  & 195.78 & \text{\texttimes} & \textbf{\checkmark} & 2 & 2 & 0.825±0.071 & 2.6±0.4 \\
        NestedMorph  & 191.48 & \textbf{\checkmark} & \text{\texttimes}   & 2 & 2 & 0.832±0.069 & 2.7±0.3 \\
        \textbf{NestedMorph}  & \textbf{202.41} & \textbf{\checkmark} & \textbf{\checkmark} & \textbf{2} & \textbf{2} & \textbf{0.840±0.072} & \textbf{2.4±0.3} \\
        NestedMorph  & 212.85 & \textbf{\checkmark} & \textbf{\checkmark} & 0 & 4 & 0.835±0.068 & 2.5±0.3 \\
        NestedMorph  & 210.63 & \textbf{\checkmark} & \textbf{\checkmark} & 1 & 3 & 0.828±0.070 & 2.8±0.4 \\
        NestedMorph  & 199.47 & \textbf{\checkmark} & \textbf{\checkmark} & 3 & 1 & 0.831±0.079 & 2.7±0.4 \\
        NestedMorph  & 196.95 & \textbf{\checkmark} & \textbf{\checkmark} & 4 & 0 & 0.838±0.075 & 2.6±0.4 \\
        \bottomrule
    \end{tabular}
    \caption{Ablation Study for NestedMorph Model. EA: Efficient Attention, CA: Channel Attention.}
    \label{tab:ablation_attention}
\end{table*}

Using a batch size of 8 yielded superior results, with a higher SSIM score and a lower HD95 compared to a batch size of 2, indicating that a larger batch size may help the model better generalize, leading to accurate segmentations. Similarly, increasing the number of heads from 1 to 8 also resulted in better performance, as seen in the higher SSIM and lower HD95. This suggests that using more attention heads can enhance the model's ability to capture complex features, thereby improving segmentation accuracy.

When comparing patch sizes, the smaller 3x3x3 patch size outperformed the larger 6x6x6 configuration, with optimal SSIM and HD95 scores. This result implies that finer patches allow the model to focus on detailed aspects of the image, leading to better segmentation performance. On the other hand, using four layers produced a slightly better SSIM and a marginally lower HD95 than using eight layers. This suggests that while deeper models are generally thought to perform better, there may be diminishing returns, and in some cases, a simpler model with fewer layers might be suitable.

\subsubsection{Attention Mechanism and Decoder Blocks}
\label{sec:attention}

We assessed the performance of Efficient Attention and Channel Attention both individually and in combination. From Table \ref{tab:ablation_attention}, it can be seen that while Channel Attention alone achieved an SSIM of 0.825, incorporating Efficient Attention alone led to a slight improvement with an SSIM of 0.832. However, the combination of both attention mechanisms resulted in the highest SSIM of 0.840 and the lowest HD95 of 2.4. This demonstrates that combining Efficient and Channel Attention optimizes performance by enhancing feature processing and computational efficiency.

We also varied the number of DAE-Former and LKA blocks in the decoder. The optimal performance was observed with 2 DAE-Former and 2 LKA blocks, achieving an SSIM of 0.840 and an HD95 of 2.4. This configuration balances feature relationship capture with contextual information. In contrast, other configurations with different combinations of DAE-Former and LKA blocks showed lower performance metrics. These results highlight the importance of balancing the number of blocks to achieve the best registration results. Overall, the combination of Efficient and Channel Attention mechanisms with 2 DAE-Former and 2 LKA blocks provides the most effective performance for the NestedMorph model.

\section{Discussion and Conclusion}

In this study, we introduced NestedMorph, a novel Nested Attention Fusion network for deformable image registration, specifically targeting the alignment of T1-weighted MRI and dMRI data. The architecture integrates high-resolution spatial details from the encoder with semantic-rich features from the decoder using a multi-scale fusion framework. This design effectively captures both fine-grained local details and long-range global dependencies, essential for accurate deformation field estimation in medical images. Our unsupervised method was rigorously evaluated against leading CNN-based and Transformer-based approaches on the HCP dataset for intra-subject registration. Results show that NestedMorph outperforms these methods, particularly in alignment accuracy. The superior performance is due to the dual attention mechanisms in the encoder, which capture both local and global dependencies, providing a balanced extraction of features compared to traditional CNN and Transformer-based models like VoxelMorph, MIDIR, TransMorph, and ViT-V-Net.

While NestedMorph excels in accuracy, it does present challenges, notably the increased computational complexity from its dual attention mechanisms and multi-scale fusion framework. These advancements, while enhancing performance, also impose higher computational demands, which may be a barrier in resource-constrained environments. Furthermore, although the model has shown promise in intra-subject registration, its effectiveness in inter-subject or cross-modality tasks remains to be explored. Future research should focus on optimizing NestedMorph for broader applications and practical clinical deployment. In conclusion, NestedMorph represents a significant leap forward in deformable medical image registration, offering unparalleled alignment of T1w and dMRI data. This study not only underscores NestedMorph's potential for transforming clinical practice but also paves the way for future advancements in deformable image registration.

{\small
\bibliographystyle{ieee_fullname}
\bibliography{egbib}
}

\end{document}


\section*{Supplementary Material}
\label{sec:supplementary}
The supplementary material provides a detailed examination of various aspects of the brain MRI registration methods and model components. \ref{sec:appendix_a} presents training curves for multiple MRI registration models, showcasing their performance in terms of SSIM scores and loss values across epochs. This analysis reveals that NestedMorph and TransMorph are the top performers, while VoxelMorph and MIDIR lag behind. \ref{sec:appendix_b} offers a comprehensive breakdown of the parameters in the NestedMorph model, highlighting a significant concentration in the Encoder and DAE-Former components. \ref{sec:appendix_c} includes further visualizations comparing eight registration techniques, illustrating their registered images, binary masks, deformation fields, and vector grids, with NestedMorph achieving the highest similarity score. Together, these appendices underscore the effectiveness of NestedMorph and TransMorph, detail the parameter distribution within NestedMorph, and provide a visual comparison of registration techniques to highlight their performance and alignment capabilities.

\begin{figure*}[hbp]
    \centering
    \includegraphics[scale=0.45]{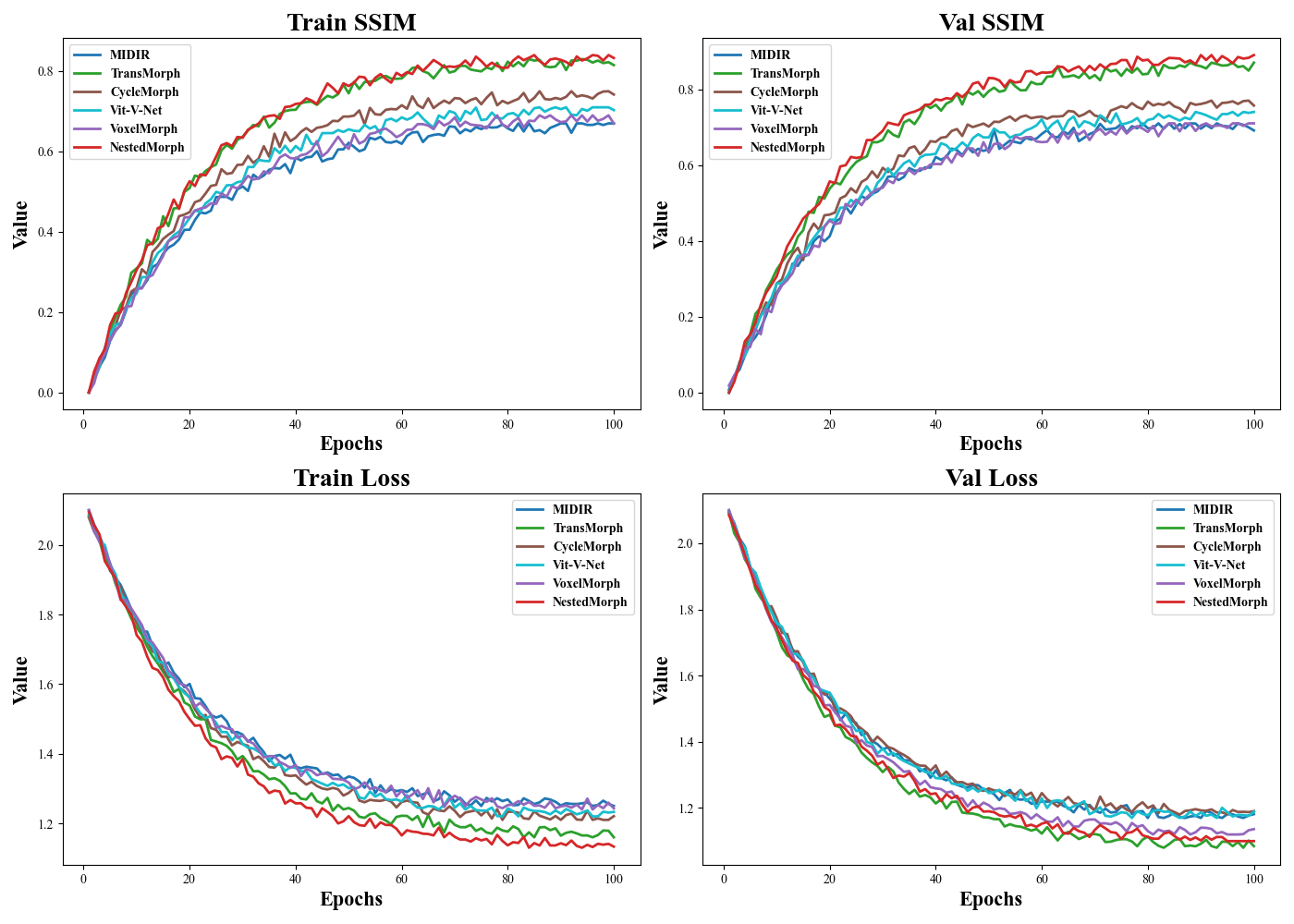}
    \caption{\textbf{Training curves for MRI registration models, showing SSIM scores and loss values.} NestedMorph and TransMorph perform best, while VoxelMorph and MIDIR lag.}
    \label{fig:training_curves}
  \hfill
\end{figure*}

\appendix
\renewcommand\thesubsection{Appendix A}

\subsection{Comparison of Training Curves Across MRI Registration Models}
\label{sec:appendix_a}

Figure \ref{fig:training_curves} shows the training curves for MRI registration models, including NestedMorph, TransMorph, CycleMorph, Vit-V-Net, VoxelMorph, and MIDIR, revealing significant performance differences. NestedMorph and TransMorph consistently outperform the others, achieving higher SSIM scores and lower loss values across training and validation phases, highlighting their superior learning efficiency and generalization capabilities. In contrast, VoxelMorph and MIDIR struggle with slower convergence and higher residual losses, reflecting difficulties in handling more complex tasks. CycleMorph and Vit-V-Net exhibit steady improvement, but their final SSIM scores and loss values suggest limitations in addressing intricate registration challenges. VoxelMorph and MIDIR further lag behind, demonstrating inefficiencies in capturing the complexity of medical images. Overall, NestedMorph and TransMorph emerge as the most effective models for accurate and robust brain MRI registration.

\appendix
\renewcommand\thesubsection{Appendix B}

\subsection{Analysis of Parameters in NestedMorph}
\label{sec:appendix_b}

\begin{table}[ht]
\centering
\begin{tabular}{@{}lcc@{}}
\toprule
\textbf{Module} & & \textbf{Params (M)} \\
\midrule
Encoder          & & 86.66 \\
\midrule
Decoder \\
\hspace{1em}DAE-Former 1 & & 87.59 \\
\hspace{1em}DAE-Former 2 & & 22.04 \\
\hspace{1em}LKA-Former 1 & & 4.05 \\
\hspace{1em}LKA-Former 2 & & 2.08 \\
\midrule
\textbf{Total} & & \textbf{202.41} \\
\bottomrule
\end{tabular}
\caption{Parameter count for each module in the model.}
\label{tab:module_params}
\end{table}

From Table \ref{tab:module_params}, the parameter distribution in NestedMorph reveals a significant focus on the Encoder and the DAE-Former components. The Encoder has the largest share, with 86.66 million parameters, reflecting its crucial role in feature extraction and transformation. The DAE-Former 1 contributes most of the remaining parameters (87.59 million), indicating its substantial role in the model's capability to perform denoising and feature refinement. The DAE-Former 2, LKA-Former 1, and LKA-Former 2 components, while still important, have comparatively smaller parameter counts, at 22.04 million, 4.05 million, and 2.08 million respectively. Overall, this distribution highlights the model's emphasis on robust feature encoding and transformation, with efficient processing managed by the more compact components.

\begin{figure*}
    \centering
    \includegraphics[scale=0.065]{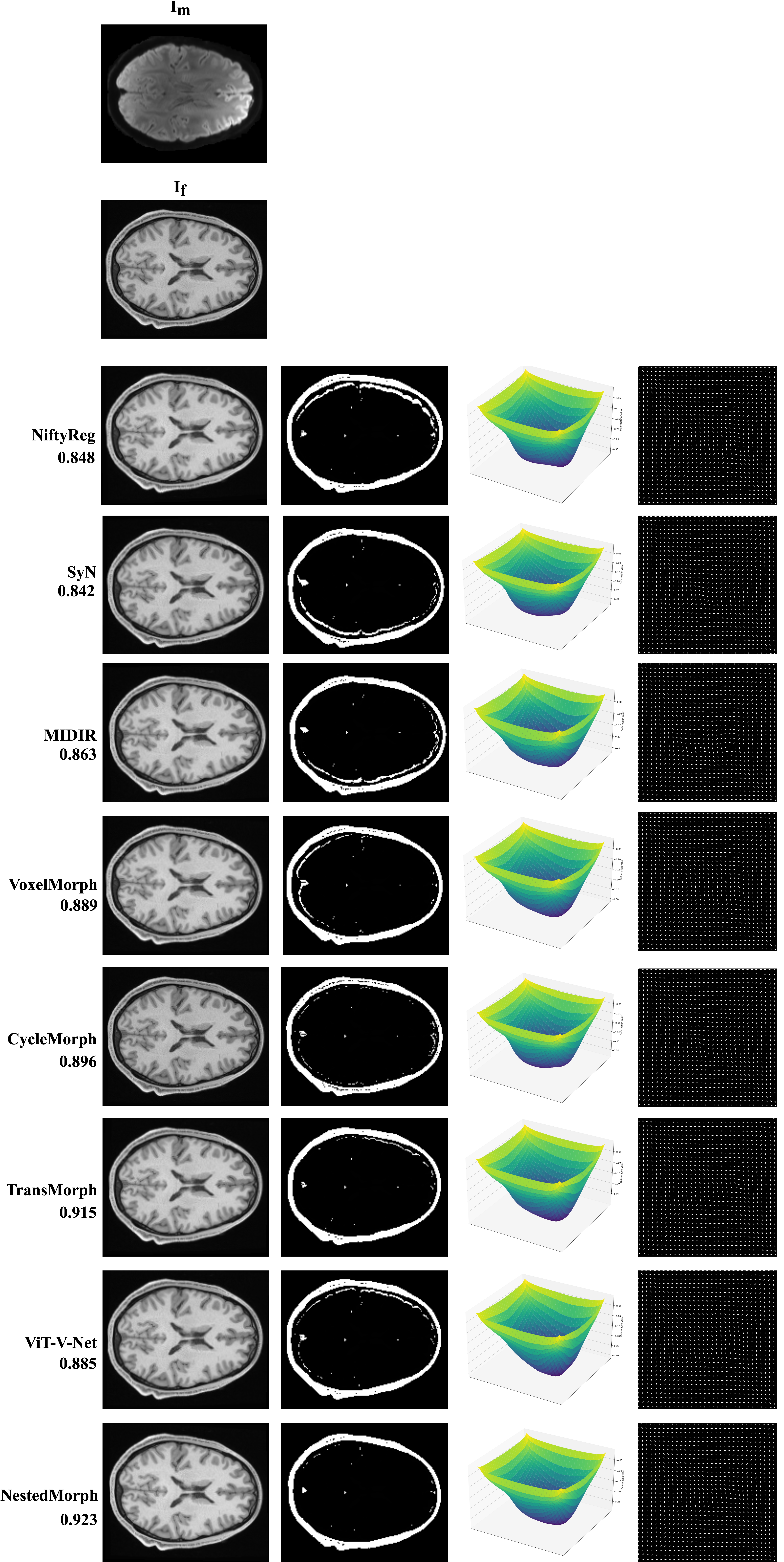}
    \caption{\textbf{Visual comparison of brain MRI registration methods. Top: Moving and fixed images.} Below: Results from eight techniques (NiftyReg, SyN, MIDIR, VoxelMorph, CycleMorph, TransMorph, VIT-V-Net, NestedMorph), with NestedMorph showing the highest similarity score (0.923).}
    \label{fig:morediagram}
  \hfill
\end{figure*}

\appendix
\renewcommand\thesubsection{Appendix C}

\subsection{Further Visualization Results}
\label{sec:appendix_c}

Figure \ref{fig:morediagram} presents a comprehensive comparison of various state-of-the-art brain MRI registration methods. The top row displays the moving (Im) and fixed (If) reference images, providing a baseline for subsequent analyses. Below, eight different registration techniques are evaluated: NiftyReg, SyN, MIDIR, VoxelMorph, CycleMorph, TransMorph, VIT-V-Net, and NestedMorph. Each method is represented by a row containing four key visualizations: the registered brain image, a binary mask highlighting the brain's contour, a 3D surface plot illustrating the deformation field, and a vector field grid depicting the transformation. Accompanying each method is a numerical value, likely representing a similarity or accuracy metric, with higher values potentially indicating superior performance. The registered images demonstrate subtle variations in alignment and detail preservation, while the deformation fields and vector grids offer insights into the underlying transformation mechanisms. Notably, NestedMorph achieves the highest score (0.923), suggesting superior performance in this comparative analysis.